\documentclass[journal]{IEEEtran}

\ifCLASSINFOpdf
\else
   \usepackage[dvips]{graphicx}
\fi
\usepackage{url}
\usepackage{graphicx}

\usepackage{color}
\usepackage{balance}
\usepackage{bm}
\usepackage{amsmath}
\usepackage{amssymb}
\usepackage{lipsum}
\usepackage{algorithm}
\usepackage{algorithmic}
\usepackage{multirow}
\usepackage{booktabs}
\usepackage{array}
\usepackage{lipsum}
\usepackage{enumerate}
\usepackage{stfloats}
\usepackage{subfigure}
\usepackage{cases}
\usepackage{diagbox}
\usepackage{cite}
\usepackage{enumitem}
\usepackage{balance}
\usepackage{geometry}
\usepackage{hyperref} 
\allowdisplaybreaks[4]
\begin{document}

\title{Sensing-Aware Transmit Waveform/Receive Filter Design for OFDM-MBS Systems}

\author{Xinghe Li, Kainan Cheng, Hongzhi Guo, \IEEEmembership{Student Member, IEEE},\\ Huiyong Li, and Ziyang Cheng, \IEEEmembership{Senior Member, IEEE}
\thanks{This work was supported in part by the National Natural Science Foundation of China under Grant 62371096, Grant 62231006; and in part by the Sichuan Science and Technology Program under 2023NSFSC1385.}
\thanks{X. Li, K. Cheng, H. Guo, H. Li and Z. Cheng are with the School of Information and Communication Engineering, University of Electronic Science and Technology of China, Chengdu 611731, China. (email: xinghe\_lee@126.com, 202452012031@std.uestc.edu.cn, hongz\_guo@126.com, hyli@uestc.edu.cn, zycheng@uestc.edu.cn).}}

\markboth{Journal of \LaTeX\ Class Files, Vol. 14, No. 8, August 2015}
{Shell \MakeLowercase{\textit{et al.}}: Bare Demo of IEEEtran.cls for IEEE Journals}
\maketitle

\begin{abstract}

In this letter, we study the problem of cooperative sensing design for an orthogonal frequency division multiplexing (OFDM) multiple base stations (MBS) system.
We consider a practical scenario where the base stations (BSs) exploit certain subcarriers to realize a sensing function.
Since the high sidelobe level (SLL) of OFDM waveforms degrades radar detection for weak targets, and the cross-correlation generated by other BSs further exacerbates detection performance, we devise a joint design scheme for OFDM sequence and receive filter by minimizing the integrated sidelobe level (ISL) while satisfying mainlobe level, peak-to-average power ratio (PAPR) and spectrum allocation constraints.
To address this non-convex problem, we propose an alternating optimization (AO)-based algorithm.
Numerical simulations validate the effectiveness of the proposed method, demonstrating the superiority of SSL reduction in the MBS system over the matched filtering method. 
\end{abstract}


\IEEEpeerreviewmaketitle

\section{Introduction}

\IEEEPARstart{I}{n} the next-generation wireless communication systems, in addition to providing communication services, sensing is also required in some emerging scenarios\cite{dong2024sensing}. Deploying multiple base stations (MBS) is a promising strategy to improve spatial diversity and enable cooperative sensing \cite{deng2004polyphase}.
Specifically, the deployment of MBS at distinct spatial locations allows the system to simultaneously observe a common target from different angles. Such spatial diversity provides richer measurement information, leading to enhanced sensing performance through spatial multiplexing and geometric diversity gains \cite{intro}.
However, employing MSE introduces hardware challenges, particularly in managing the trade-off between spectral efficiency and power consumption \cite{he2024_15,Roos_Bechter_Knill_Schweizer_Waldschmidt_2019,Wei_Wu_Mishra_Shankar_2022}. 
To address these issues, orthogonal frequency division multiplexing (OFDM) waveforms, which are known for their high spectral efficiency and seamless compatibility with existing communication standards, have been considered as a practical candidate for sensing functionalities in MBS systems \cite{Roos_Bechter_Knill_Schweizer_Waldschmidt_2019}.

Compared with traditional radar waveforms such as linear frequency modulation (LFM), OFDM waveforms exhibit several advantages, including higher range resolution \cite{nuss2020frequency9}, improved Doppler tolerance \cite{nuss2018limitations10}, and greater waveform diversity \cite{han2023sub12}, making them well-suited for radar sensing applications \cite{lin2019non11}.
However, due to the non-ideal autocorrelation and cross-correlation properties of OFDM sequences, high sidelobe levels (SLLs) can arise in range estimation, which degrade sensing performance \cite{Sturm_Pancera_Zwick_Wiesbeck_2009,Bai_Hu_Song_2017,Shi_Wang_He_Cheng_2019}.
Facing these challenges, some waveform optimization techniques have been proposed \cite{cui2017constant,cheng2019nonlinear,xu2025cooperative}.
For instance, authors in \cite{cui2017constant} introduced a block coordinate descent (BCD)-based approach to minimize the weighted integrated sidelobe level (WISL). 
Building on this, a nonlinear alternating direction method of multipliers (Nonlinear-ADMM) was proposed in \cite{cheng2019nonlinear}, which directly optimized the WISL, achieving better sidelobe suppression. 
Furthermore, \cite{xu2025cooperative} improves upon these methods by assigning different weights to autocorrelation and cross-correlation terms under the constraints of the OFDM time-frequency structure, leading to further reduction in SLLs.

In addition to the aforementioned sidelobe issues, OFDM sequences suffer from a high peak-to-average power ratio (PAPR) \cite{Hu_Masouros_Liu_Nissel_2022,Cheng_He_Liao_Fang_2018}, which not only reduces transmitter efficiency but also increases hardware complexity and energy consumption \cite{wang2022exploiting,Sen_2014}. 
To alleviate this issue, authors in \cite{lellouch2015convex} proposed an optimization framework to reduce the PAPR of OFDM radar waveforms while maintaining desirable sensing properties.
Moreover, authors in \cite{barneto2019full} implemented PAPR reduction techniques in the transmitter chain and employed OFDM symbol windowing. 
Another work in \cite{Tsai_Chung_Shiu_2011} proposed a Gerchberg-Saxton (GS) algorithm to adjust the phases of the transform-domain sequence, which helped to lower the PAPR while maintaining the optimal autocorrelation property.
Although the work \cite{lellouch2015convex,barneto2019full,Tsai_Chung_Shiu_2011} achieved noticeable sidelobe level (SLL) reduction, they have two key limitations:
1) they are primarily designed for single-node radar systems and do not address the spatial coordination and signal diversity challenges intrinsic to distributed base station (BS) deployment;
2) the resulting waveform performance is still insufficient for the reliable detection of weak targets in dense urban environments, due to residual interference arising from the imperfect autocorrelation and cross-correlation properties of OFDM waveforms.

Motivated by the facts discussed above, this letter makes the following contributions. 
\textit{First,} we focus on the design of low-correlation sidelobe OFDM waveforms in the context of MBS systems, taking into account both the transmit OFDM sequence and the receive filter.
\textit{Second,} based on the time-frequency structure, we devised a joint design problem by minimizing the integrated sidelobe level (ISL) subject to spectral occupancy, PAPR, and main lobe constraints.
Then, we propose an alternating optimization (AO)-based algorithm to solve this non-convex problem.
Simulation results demonstrate the effectiveness of the proposed design, achieving superior low-sidelobe performance for MBS-based joint sensing applications.

\section{Signal Model and Problem Formulation}

We consider an OFDM-based MBS system, where $M$ BSs aim to reuse a subset of the available $N$ subcarriers for radar sensing purposes while simultaneously maintaining their original communication service.
Specifically, $N$ subcarriers are divided into two complementary groups, $\mathcal N$ and $\bar{\mathcal N}$.
They collaboratively exploit a subset $\mathcal N$ to cooperatively probe a weak target in the environment.

\subsection{Signal Model}

\subsubsection{Transmit Signal}
Suppose ${\bf{s}} = {[{{\mathbf{s}}}_1^T,{{\mathbf{s}}}_2^T...,{{\mathbf{s}}}_M^T]^T} \in {\mathbb C}^{MN}$ denotes the collection of the frequency-domain sensing signals for all the BSs, where ${\mathbf{s}_m} ={\left[ {s_{m}(1), \ldots ,{s_{m}(N)}} \right]^T}$ denotes the frequency-domain waveform of the $m$-th BS.
Since only the subcarrier subset $\mathcal N$ is exploited for the sensing function, we assume $\left| s_{m}(n) \right| = 1$ only if $n \in {\mathcal N}$.
After the discrete Fourier transformation (DFT), the time-domain signal is given by 
\begin{equation}\label{eq:1}
	\begin{aligned}
		\mathbf{x}=(\mathbf{I}_M\otimes\mathbf{F}^H)\mathbf{s},
	\end{aligned}
\end{equation}
where $\mathbf{F}\in\mathbb{C}^{N\times N}$ denotes a DFT matrix with deflation to $1/N$ of the original one.
Note that the signal \eqref{eq:1} can be expressed $\mathbf{x} = {[\mathbf{x}_1^T,\mathbf{x}_2^T...,\mathbf{x}_M^T]^T}$, where ${\mathbf{x}_m} ={\left[ {{x_{m}(1)}, \ldots ,{x_{m}(N)}} \right]^T}$ denotes the time-domain waveform of the $m$-th BS.

\subsubsection{Performance Metric}
Since the effectiveness of matched filters at the receivers is determined by autocorrelation and cross-correlation of the received signal, we introduce the ISL as the sensing performance metric \cite{cheng2019nonlinear}.
Specifically, the ISL can be defined as
\begin{equation}\label{eq:2}
\begin{aligned}
\varepsilon_{r}=&\sum_{m=1}^{M}\sum_{l=-N+1}^{N-1}\left|r_{mm}(l)\right|^2\\
&+\sum_{m_1=1}^{M}\sum_{m_2=1}^{M}\sum_{l=-N+1}^{N-1}\left|r_{{m_1}{m_2}}(l)\right|^2,
\end{aligned}
\end{equation}
where ${r_{mm}}(l)$ represents the ${l}$-th sample of the time-domain autocorrelation function of the ${m}$-th BS, and ${r_{{m_1}{m_2}}}(l)$ represents the ${l}$-th sample of the time-domain cross-correlation function of the ${m_1}$-th BS and ${m_2}$-th BS.
They can be expressed as
\begin{subequations}\label{eq:3}
	\begin{align}&{r_{mm}}(l)=\sum_{k=l+1}^{N}{x}_{m}(k){h}_{m}(k-l)={\mathbf{x}_{m}\mathbf{E}_{l}\mathbf{h}_{m}},\\
    &r_{{m_1}{m_2}}(l)=\sum_{k=l+1}^{N}{x}_{{m_1}}(k){h}_{{m_2}}(k-l)={\mathbf{x}_{{m_1}}}{\mathbf{E}_{l}}{\mathbf{h}_{m_2}},
	\end{align}
\end{subequations}
where ${{\mathbf{E}_l}\in\mathbb{R}^{N\times N}}$ denotes a shift matrix whose $(i,j)$ entry equals 1 when $i - j = l$ and 0 otherwise, and ${\mathbf{h}_m} ={\left[ {{h_{m}(1)}, \ldots ,{h_{m}(N)}} \right]^T\in\mathbb{C}^{N}}$ denotes the receive filter at the $m$-th BS.
We define ${\bf h} = [{\bf h}_1,\dots,{\bf h}_M]$ as the collection of the receiver filters for all the BSs.
Note that a lower ISL leads to better sensing performance.

\vspace{-1em}
\subsection{Problem Formulation}

In this letter, to improve the cooperative sensing performance, we aim to jointly design the transmit waveforms $\bf s$ and receiver filters $\bf h$ for the considered MBS system.
Mathematically, the design problem can be formulated as
\begin{subequations}
\begin{align}
&\mathop{\min}\limits_{\mathbf{s}, \mathbf{h}} 
\sum_{m=1}^M \sum_{l = -N+1}^{N-1} \!\!\!\!\left| {r_{mm}}(l) \right|^2 
+ \sum_{{m_1}=1}^M \sum_{{m_2}=1}^M \sum_{l = -N+1}^{N-1}\!\!\!\! \left|r_{{m_1}{m_2}}(l)\right|^2,\label{eq:P1-a}\\
&\;\;{\rm{s.t.}}  \;\;\mathbf{x}_m^H \mathbf{h}_m \geq \gamma, \forall m,\label{eq:P1-b}\\
& \qquad\; \frac{\left| {x}_m(n) \right|^2}{P} \leq \eta, \forall m, n,\label{eq:P1-c}\\
& \qquad\; \left| s_{m}(n) \right| = 
      \begin{cases} 
        1, & n \in \mathcal{N} \\
        0, & n \in \bar{\mathcal{N}}
      \end{cases} ,\forall m,n,\label{eq:P1-d}
\end{align}
\label{eq:P1}%
\end{subequations}
where \eqref{eq:P1-b} denotes the mainlobe gain requirement with threshold $\gamma$, \eqref{eq:P1-c} denotes the PAPR constraint with threshold $\eta$ and average power $P$, and \eqref{eq:P1-d} denotes the spectrum allocation constraint.

Since \eqref{eq:P1} is an non-convex problem, involving a close coupling relationship between primal variables in the objective function and constraints, as well as a constant modulus constraint, it is hard to tackle this problem using conventional methods.
To this end, we solve this problem in the next section.

\section{Solution to Problem \eqref{eq:P1}}

In this section, we first transform the original problem into a more tractable form and then derive solutions to the problem.
Finally, we conclude the proposed algorithm and analyze its complexity.

\subsection{Problem Transformation}

To simplify the objective function, \eqref{eq:3} can be substituted into \eqref{eq:P1-a}, leading to the reformulated problem
\begin{subequations}
\begin{align}
&\mathop{\min}\limits_{\mathbf{s},\mathbf{h}} \quad 
\sum_{l = -N+1}^{N-1} \left\| \mathbf{X}^H \mathbf{E}_l \mathbf{H} \right\|_F^2 
- \sum_{m=1}^M \sum_{l = -i}^i \left| \mathbf{x}^H \mathbf{\tilde{E}}_{lm} \mathbf{h} \right|^2, \label{eq:P2-a}\\
& \;\;{\rm{s.t.}} \;\;\left| \mathbf{x}^H \mathbf{\bar{E}}_m \mathbf{h} \right| \geq \gamma, \forall m \label{eq:P2-b}\\
& \qquad \;\;\eqref{eq:P1-c} \;{\rm and} \;\eqref{eq:P1-d},
\end{align}
\label{eq:P2}%
\end{subequations}
where we define ${\mathbf{\tilde{E}}}_{lm} =\operatorname{Bdiag}\bigl( {\bf{0}}_N,\ldots, {\mathbf{E}_{l}}, \ldots,{\bf{0}}_N \bigr)$ with the ${m}$-th diagonal block $\mathbf{E}_{l}$, and define ${\mathbf{\bar{E}}}_{m} = {\rm Bdiag}({\bf 0}_N,\dots, {\bf I}_N,\dots,{\bf 0}_N)$ with the ${m}$-th diagonal block ${{\bf{I}}_N}$.  
Besides, $\mathbf{X}=[\mathbf{x}_{1},\mathbf{x}_{2},...,\mathbf{x}_{M}]\in\mathbb{C}^{N\times M}$ and $\mathbf{H}=[\mathbf{h}_{1},\mathbf{h}_{2},...,\mathbf{h}_{M}]\in\mathbb{C}^{N\times M}$ . 
Note that the first term in \eqref{eq:P2-a} represents the total energy of the MBS system, and the second term represents the energy of the mainlobe region, where $i$ denotes the number of sampling points symmetrical about the mainlobe center.

Then, to decouple the primal variables in the objective function and constraints, we introduce auxiliary variables ${\bf y} \in {\mathbb C}^{MN}$ and $\{z_m\}$ and transform the problem \eqref{eq:P2} into 
\begin{subequations}
\begin{align}
&\mathop{\min}\limits_{\mathbf{s},\mathbf{h}, {\bf y},\{z_m\}} 
\sum_{l = -N+1}^{N-1}\!\!\!\! \left\| \mathbf{Y}^H \mathbf{E}_l \mathbf{H} \right\|_F^2 
- \sum_{m=1}^M \sum_{l = -i}^i \left| \mathbf{y}^H \mathbf{\tilde{E}}_{lm} \mathbf{h} \right|^2, \label{eq:P3-a}\\
& \;\;\quad{\rm{s.t.}} \;\;\quad\left| z_m\right| \geq \gamma,  \forall m, \label{eq:P3-b}\\
&\;\; \qquad \qquad\frac{\left| {y}_m(n) \right|^2}{P} \leq \eta, \forall m,n, \label{eq:P3-c}\\
&\;\;\qquad\qquad \left| s_{m}(n) \right| = 
      \begin{cases} 
        1, & n \in \mathcal{N} \\
        0, & n \in \bar{\mathcal{N}}
      \end{cases},\forall m,n,\label{eq:P3-d}\\
      & \;\;\qquad\qquad \mathbf{y}=(\mathbf{I}_M\otimes\mathbf{F}^H)\mathbf{s},\label{eq:P3-e}\\ 
      & \;\;\qquad\qquad{z_m} = \mathbf{y}^H \mathbf{\bar{E}}_m \mathbf{h}, \forall m,\label{eq:P3-f}
\end{align}
\label{eq:P3}%
\end{subequations}
where we define $\mathbf{Y}=[\mathbf{y}_{1},\mathbf{y}_{2},...,\mathbf{y}_{M}]\in\mathbb{C}^{N\times M}$.
Penalizing the equality \eqref{eq:P3-e} and \eqref{eq:P3-f} into the objective function, the problem \eqref{eq:P3} can be reformulated into an augmented Lagrange minimization problem, which can be expressed as
\begin{equation}\label{eq:P4}
    \begin{aligned}
    &\mathop{\min}\limits_{{\mathbf{y},\mathbf{h},{\bf{s}},\{z_m\},{\mathbf{u}}, {\{v_m\}} } } {\mathcal{L}}\left( {\mathbf{y},\mathbf{h},{\bf{s}},\{z_m\},{\mathbf{u}}, {\{v_m\}} } \right),\\
    &\qquad\quad{\rm s.t.}\qquad \;\;\eqref{eq:P3-b}-\eqref{eq:P3-f},
    \end{aligned}
\end{equation}
where ${\mathcal{L}}\left( {\mathbf{y},\mathbf{h},{\bf{s}},\{z_m\},{\mathbf{u}}, {\{v_m\}} } \right) = \sum\nolimits_{l =  - N + 1}^{N - 1} {\left\| {{{\bf{Y}}^H}{{\bf{E}}_l}{\bf{H}}} \right\|_F^2}  - \sum\nolimits_{m = 1}^M {\sum\nolimits_{l =  - i}^i {{{\left| {{{\bf{y}}^H}{{{\bf{\tilde E}}}_{lm}}{\bf{h}}} \right|}^2}} } + \frac{{{\rho _u}}}{2} {\left\| {\mathbf{y}-(\mathbf{I}_M\otimes\mathbf{F}^H)\mathbf{s} + {\bf{u}}} \right\|_2^2} + \frac{{{\rho _v}}}{2}\sum\nolimits_{m = 1}^M {\left\| {{z_m} - {\mathbf{y}^H}{{{\bf{\bar E}}}_m}{\mathbf{h}} + {v_m}} \right\|_2^2}$, with the penalty parameters ${\rho _u}$, ${\rho _v}>0$ and corresponding dual variables ${\mathbf{u}}, {\{v_m\}}$.

\vspace{-1em}
\subsection{Alterating Optimization of Problem \eqref{eq:P4}}
Solutions to the problem \eqref{eq:P4} can be obtained alternatively by solving the following subproblems. 

\subsubsection{Update $\mathbf{y}$}
With other variables fixed, ${\mathbf{y}}$ can be obtained by solving the subproblem
\begin{subequations}
\begin{align}
&\mathop{\min}\limits_{\mathbf{y}} 
\sum_{l = -N+1}^{N-1} \left\| \mathbf{Y}^H \mathbf{E}_l \mathbf{H} \right\|_F^2 
- \sum_{m=1}^M \sum_{l = -i}^i \left| \mathbf{y}^H \mathbf{\tilde{E}}_{lm} \mathbf{h} \right|^2 \nonumber\\
&\qquad\quad  + \frac{\rho_u}{2} \sum_{m=1}^M \left\|\mathbf{y}-(\mathbf{I}_M\otimes\mathbf{F}^H)\mathbf{s} + {\bf{u}} \right\|_2^2 \nonumber\\
& \qquad\quad+ \frac{\rho_v}{2} \sum_{m=1}^M \left\| z_m - {\mathbf{y}}^H\mathbf{\bar{E}}_m \mathbf{h} + v_m \right\|_2^2 ,\\
&\;\; {\rm{s.t.}} \quad\left| y_m(n) \right|^2 \leq \eta P, \forall m,n,\label{eq:8-b}
\end{align}\label{eq:8}%
\end{subequations}
which is a convex problem.
By leveraging the Karush-Kuhn-Tucker (KKT) conditions, we have 
\begin{equation}
{\mathbf{y}} = {{\mathbf{\Phi }}^{ - 1}}{\mathbf{b}},
\end{equation}
with defining
\begin{equation}
    \begin{aligned}
        {\mathbf{\Phi }} =& 2\sum\limits_{l =  - N + 1}^{N - 1} {{{\mathbf{D}}_l}{\mathbf{H}}}  - 2\sum\limits_{m = 1}^M {\sum\limits_{l =  - i}^i {{{{\mathbf{\tilde E}}}_{lm}}{\mathbf{h}}{{\mathbf{h}}^H}{\mathbf{\tilde E}}_{lm}^H} }  \\
        &+ {\rho _u}{\mathbf{I}} + {\rho _v}\sum\limits_{m = 1}^M {{{{\mathbf{\bar E}}}_m}{\mathbf{h}}{{\mathbf{h}}^H}{\mathbf{\bar E}}_m^H} ,\\
        {\bf b} =& {{\rho _u}\left( {\left( {{{\mathbf{I}}_M} \otimes {{\mathbf{F}}^H}} \right){\mathbf{s}} - {\mathbf{u}}} \right) + {\rho _v}\sum\limits_{m = 1}^M {{{{\mathbf{\bar E}}}_m}{\mathbf{h}}{{\left( {{z_m} + {v_m}} \right)}^*}} }.
    \end{aligned}\nonumber
\end{equation}
Then, taking the constraint \eqref{eq:8-b} into consideration, $\bf y$ can be updated by\footnote{In this letter, $(\cdot)^{k+1}$ denotes the next point of $(\cdot)^{k}$, with $k$ denoting the iteration number.}
\begin{equation}\label{eq:10}
    {y_m}{(n)^{(k + 1)}} = \left\{ \begin{gathered}
  \frac{{\sqrt {\eta P} {\mkern 1mu} {y_m}{{(n)}^{(k)}}}}{{\left| {{y_m}{{(n)}^{(k)}}} \right|}},{\left| {{y_m}(n)} \right|^2} > \eta P \hfill \\
  \quad {y_m}{(n)^{(k)}}\quad \;,{\rm{otherwise}} \hfill \\ 
\end{gathered}  \right..
\end{equation}

\subsubsection{Update $\mathbf{h}$}
With other variables fixed, ${\bf{h}}$ can be updated by solving the subproblem
\begin{equation}
\begin{aligned}
&\mathop{\min}\limits_{\mathbf{h}} 
\sum_{l = -N+1}^{N-1} \left\| \mathbf{Y}^H \mathbf{E}_l \mathbf{H} \right\|_F^2 
- \sum_{m=1}^M \sum_{l = -i}^i \left| \mathbf{y}^H \mathbf{\tilde{E}}_{lm} \mathbf{h} \right|^2 \\
& \qquad\quad+ \frac{\rho_v}{2} \sum_{m=1}^M \left\| z_m - {\mathbf{y}}^H\mathbf{\bar{E}}_m \mathbf{h} + v_m \right\|_2^2,
\end{aligned}\label{eq:11}
\end{equation}
which is a convex problem without any constraints.
To this end, the solution to $\bf h$ can be obtained by
\begin{equation}\label{eq:12}
{{\mathbf{h}}^{(k + 1)}} = {{\mathbf{\Psi }}^{ - 1}}{\mathbf{g}},
\end{equation}
with defining
\begin{equation}
    \begin{aligned}
        {\mathbf{\Psi }} =& 2\sum\limits_{l =  - N + 1}^{N - 1} {{{\mathbf{D}}_l}{\mathbf{Y}}}  - 2\sum\limits_{m = 1}^M {\sum\limits_{l =  - i}^i {{\mathbf{\tilde E}}_{lm}^H{\mathbf{y}}{{\mathbf{y}}^H}{{{\mathbf{\tilde E}}}_{lm}}} }  \\
        &+ {\rho _v}\sum\limits_{m = 1}^M {{\mathbf{\bar E}}_m^H{\mathbf{y}}{{\mathbf{y}}^H}{{{\mathbf{\bar E}}}_m}} ,\\
        {\mathbf{g}} = &{\rho _v}\sum\limits_{m = 1}^M {{\mathbf{\bar E}}_m^H} {\mathbf{y}}{\left( {{z_m} + {v_m}} \right)^*}.
    \end{aligned}\nonumber
\end{equation}

\subsubsection{Update $\mathbf{s}$}

With other variables fixed, $\bf s$ can be obtained by solving the subproblem
\begin{subequations}
\begin{align}
\mathop{\min}\limits_{\mathbf{s}} & \quad {\left\| {\mathbf{y}-(\mathbf{I}_M\otimes\mathbf{F}^H)\mathbf{s} + {\bf{u}}} \right\|_2^2{\kern 1pt} } ,\\
\text{s.t.} & \quad \left| s_{m}(n) \right| = 
      \begin{cases} 
        1, & n \in \mathcal{N} \\
        0, & n \in\bar{\mathcal{N}}
      \end{cases}.
\end{align}\label{eq:13}%
\end{subequations}
To tackle this, we transform \eqref{eq:13} into an equivalently problem:
\begin{equation}
\begin{aligned}
\mathop{\min}\limits_{\mathbf{s}} & \quad 
{ - 2{\text{Re}}\left\{ {{{\left( {{\mathbf{y}} + {\mathbf{u}}} \right)}^H}\left( {{{\mathbf{I}}_M} \otimes {{\mathbf{F}}^H}} \right){\mathbf{s}}} \right\}}  + {\rm Const.}, \\
{\rm{s.t.}} & \quad \left| s_{m}(n) \right| = 
      \begin{cases} 
        1, & n \in \mathcal{N} \\
        0, & n \in\bar{\mathcal{N}}
      \end{cases} ,\forall m,n,
\end{aligned}
\end{equation}
whose closed-form solution can be derived as
\begin{equation}\label{eq:15}
\begin{aligned}
 s_{m}(n) = 
      \begin{cases} 
        e^{\jmath\arg(f_m(n))}, & n \in \mathcal{N} \\
        0, & n \in\bar{\mathcal{N}}
      \end{cases} , \forall m,n,\\
\end{aligned}
\end{equation}
with defining $\mathbf{f}=(\mathbf{I}_{M}\otimes\mathbf{F}^H)^{H}(\mathbf{y+u})\in\mathbb{C}^{MN}$, and $f_m(n)$ denoting the $n$-th item of the $m$-th segment.

\subsubsection{Update $\{{z_m}\}$}
With other variables fixed, $\{z_m\}$ can be solved by solving the subproblem 
\begin{subequations}
    \begin{align}
        &\mathop{\min}\limits_{\{z_m\}} \sum_{m=1}^M \left\| z_m - {\mathbf{y}}^H\mathbf{\bar{E}}_m \mathbf{h} + v_m \right\|_2^2 ,\\
        &\;\;{\rm s.t.} \;\left| z_m\right| \geq \gamma,  \forall m,\label{eq:16-b}
    \end{align}\label{eq:16}%
\end{subequations}
whose closed-form solution can be easily obtained by
\begin{equation}\label{eq:17}
	\begin{aligned}
{z_m} = { {\mathbf{y}}^H}{{\bf{\bar E}}_m}{{\mathbf{h}}} - v_m.
         \end{aligned}
\end{equation}
Then, forcing \eqref{eq:17} into the feasible region \eqref{eq:16-b}, ${{z_m}}$ is updated by
\begin{equation}\label{eq:18}
{z_m}^{(k + 1)} = \left\{ \begin{gathered}
  \quad {z_m}\;\;\;,\left| {{z_m}} \right| \geqslant \gamma  \hfill \\
  \frac{{\sqrt \gamma  {\mkern 1mu} {z_m}}}{{\left| {{z_m}} \right|}},{\text{otherwise}} \hfill \\ 
\end{gathered}  \right..
\end{equation}

\subsubsection{Update ${\bf u},{\{z_m\}}$}
With other variables fixed, dual variables can be updated by 
\begin{equation}
    \begin{aligned}
        &{{{\mathbf{u}}}^{\left( {k + 1} \right)}} = {{{\mathbf{u}}}^{\left( k \right)}} + \mathbf{y}^{(k+1)}  {-(\mathbf{I}_M\otimes\mathbf{F}^H)\mathbf{s}^{(k+1)}},\\
        &{v_m}^{(k + 1)} = {z_m}^{(k + 1)} - {{\mathbf{y}^{{(k + 1)}}}}{{\bf{\bar E}}_m}{{\mathbf{h}^{(k + 1)}}} + {v_m}^{(k)}.
    \end{aligned}\label{eq:19}
\end{equation}

\setlength{\textfloatsep}{0.5em}
\begin{algorithm}[!t]
	\caption{Proposed algorithm for the MBS system}
	\label{alg:1}
	\begin{algorithmic}[1]
		\STATE {\bf{Input:}} System parameters.
        \STATE {\bf{Initialization:} } $\left\{ {\mathbf{y}, {\bf{h}}, {\bf{s}},\{{z_m}}\} \right\}$, $k=0$.
		\WHILE{No Convergence}
        \STATE $k = k + 1$.
		\STATE Update $\mathbf{y}^{(k)}$ by solving \eqref{eq:10}.
		\STATE Update $\mathbf{h}^{(k)}$ by solving \eqref{eq:12}.
        \STATE Update ${\mathbf{s}}^{(k)}$ by solving \eqref{eq:15}.
        \STATE Update $\{{z_m}\}^{(k)}$ by solving \eqref{eq:18}.
        \STATE Update dual variables by \eqref{eq:19}.
		\ENDWHILE
		\STATE \textbf{Output:} Optimal transmit sequences $\mathbf{x}=(\mathbf{I}_M\otimes\mathbf{F}^H){\mathbf{s}}^{(k)}$ and receive filter ${\mathbf{h}} = \mathbf{h}^{(k)}$.
	\end{algorithmic}
\end{algorithm}

Based on the above derivations, we conclude the proposed algorithm in Algorithm \ref{alg:1}.
The complexity of solving for $\mathbf{x}$ and $\mathbf{h}$ is $\mathcal{O}(M^3N^3)$ due to the effect of inversion, solving for $\mathbf{s}$ is $\mathcal{O}(MN\log N)$ and solving for ${z_m}$.
To sum up, the overall complexity of the proposed algorithm is ${\mathcal{O}}(I_{\rm AO}({M^3N^3 + MN\log N}))$, where $I_{\rm AO}$ denotes the iteration number.

\section{Numerical Simulations}

In this section, we conduct numerical simulations to assess the performance of the proposed design.
It is assumed that the number of distributed BSs is ${M=2}$ and the number of OFDM subcarriers for each BS is ${N=256}$. 
We set PAPR constraint as $\eta  = 1.5$ and the index set of unavailable subcarrier sequences as $\bar{\mathcal{N}}=\{114, 105, \dots, 142\}$.

\begin{figure}[t]
	\centering
	\subfigure[]{
		\label{fig:subfig:3a}
		\includegraphics[width=0.8\linewidth]{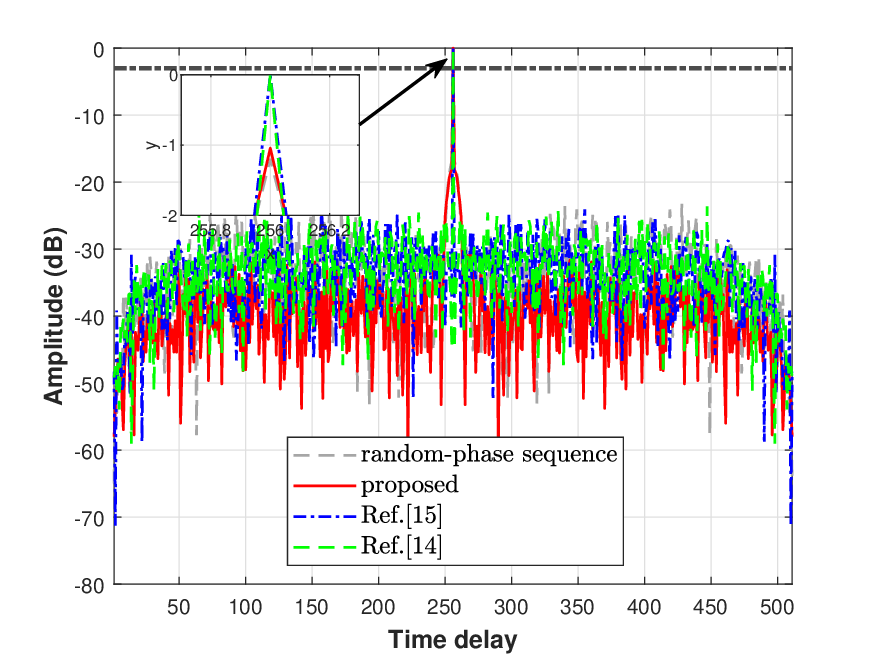}}\\[-1.25ex]  
	\subfigure[]{
		\label{fig:subfig:3b}
		\includegraphics[width=0.8\linewidth]{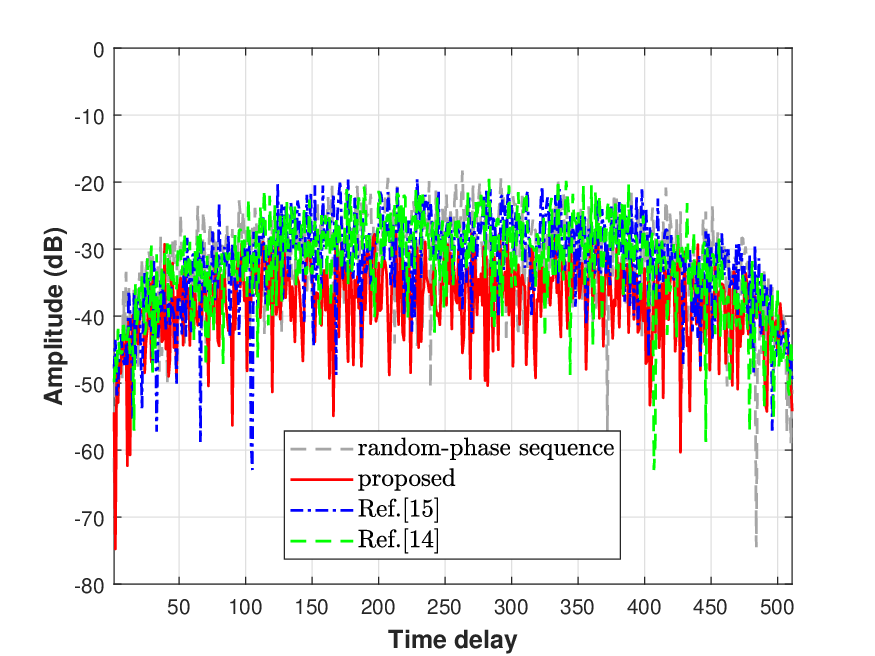}}
        \vspace{-0.5em}
	\caption{Normalized correlation sidelobe level characteristics, (a) represents the autocorrelation function and (b) represents the cross-correlation function.}
	\label{fig:pic3}
\end{figure}

Fig. \ref{fig:pic3} compares the normalized correlation sidelobe characteristics of the initial random phase, the algorithm mentioned in \cite{cui2017constant} (BCD) and \cite{cheng2019nonlinear} (Nonlinear-ADMM) which adopted the matched filtering and the proposed algorithm. Due to the spectrally limited constraint, the total energy in the time-domain is ${(N - \left| \bar{\mathcal{N}} \right|)}/{N}$. Therefore, the mainlobe will lose about 1dB. As shown in Fig. \ref{fig:subfig:3a} and \ref{fig:subfig:3b} , it is obvious that the proposed algorithm is significantly lower than the initial sequence and is about 10 dB lower than the matched filtering in terms of autocorrelation and cross-correlation. We ignore part of the mainlobe energy through the second term in \eqref{eq:P2-a}, but we still can find that the mainlobe width is not significantly broadened.


\begin{figure}[htbp]
\vspace{-1em}
	\centering
	\includegraphics[width=0.8\linewidth]{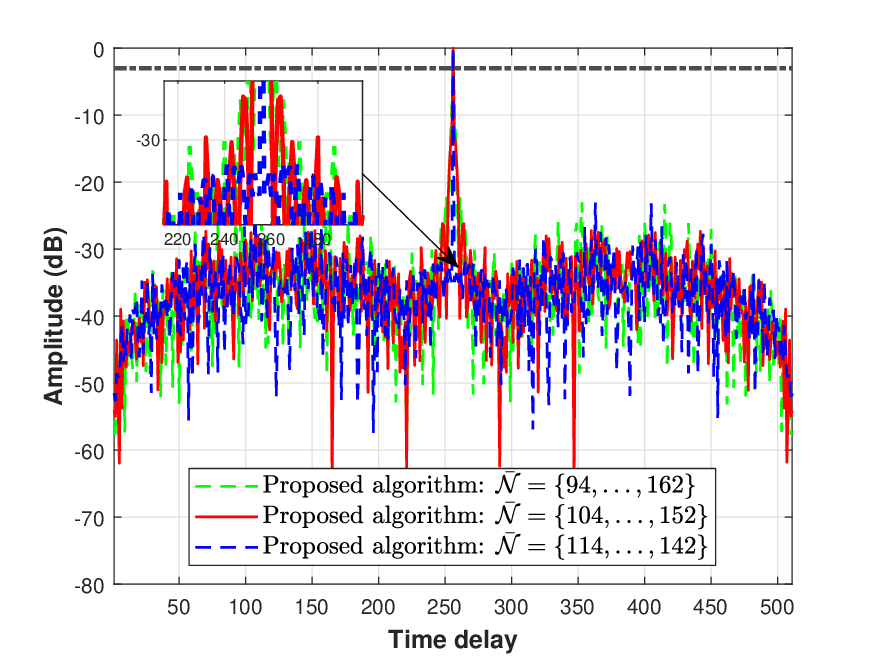}
    \vspace{-0.5em}
	\caption{ Autocorrelation of different number of unavailable subcarriers.}
	\label{fig:pic5}
\end{figure}

Fig. \ref{fig:pic5} presents the  correlation results for different numbers of unavailable subcarriers. It can be easily observed that as the number of available subcarriers increases, the SLL after the mismatched filter become lower and the mainlobe width is narrower. It is because more available subcarriers can provide higher Degrees of Freedom (DoF), which can be utilized to optimize the objective function, thus showing better SLL performance. 

\section{Conclusion}

In this letter, we have studied the joint design of sensing waveform and receive filter for an MBS system. 
The high autocorrelation and cross-correlation caused by the OFDM MBS system and MBS hardware limitations pose a challenge for radar detection. 
To address this problem and achieve a better detection performance, we formulate an ISL minimization problem with the requirements of mainlobe level, PAPR and spectrum allocation.
To solve the non-convex optimization problem, this letter proposes an AO-based algorithm. 
Simulation results demonstrate the effectiveness of the proposed design in reducing SLL for the MSB system.

\newpage
\balance
\appendices
\bibliographystyle{ieeetr}
\bibliography{IEEEabrv,stan_ref}

\end{document}